\pdfoutput=1
\documentclass[showpacs,preprintnumbers,amsmath,amssymb]{revtex4}

\usepackage{graphicx}
\usepackage{bm}

\DeclareMathOperator{\sign}{sign}

\begin{document}

\title{Existence of localised solutions in one-dimensional nonlinear lattices}

\author{Dirk Hennig}
\affiliation{Institut f\"ur Physik, Humboldt-Universit zu Berlin,
Newtonstr. 15, 12489 Berlin, Germany}
\author{J. Cuevas--Maraver \thanks{Email: jcuevas@us.es}}
\affiliation{Grupo de F\'{\i}sica No Lineal, Departamento de F\'{\i}sica Aplicada I,
Universidad de Sevilla. Escuela Polit{\'e}cnica Superior, C/ Virgen de
\'Africa, 7, E-41011 Sevilla, Spain}
\affiliation{Instituto de Matem\'aticas de la Universidad de Sevilla (IMUS). Edificio
Celestino Mutis. Avda. Reina Mercedes s/n, E-41012 Sevilla, Spain}

\date{\today}

\begin{abstract}
\noindent
We prove the existence of spatially localised and time-periodic solutions
in general nonlinear Hamiltonian Klein-Gordon lattice systems.
Like normal modes, these localised solutions
are characterised by
collective oscillations at the lattice sites with a uniform time-dependence, 
an approximation that is
valid for lattice states
near the end points of the linear spectral band. The proof of existence
uses the comparison
principle for differential equations to demonstrate that at each lattice site
 every half of the fundamental period of oscillations, contributing to
 localised solutions of the nonlinear lattice,
is sandwiched between two oscillatory states of auxiliary linear equations.
For soft (hard) on-site potentials
the  allowed frequencies of the in-phase (out-of-phase) localised periodic solutions
lie
 below (above) the lower (upper) value of the linear spectrum of phonon frequencies. 
 A numerical analysis of the solution behaviour verifies that localised solutions with frequencies 
 close to the edges of the the linear phonon spectrum exhibit (virtually) uniform time dependence.   
\end{abstract}

\maketitle

\section{Introduction}

Ever since their occurrence
was proposed in \cite{Sievers},
intrinsic localised modes (ILMs) or discrete breathers in nonlinear lattices
have  attracted continued interest, not least due to the important role they play in many
physical realms where features
of localisation in systems of coupled oscillators are involved
(for a review see \cite{Flach1} and references
therein),\cite{Aubry}-\cite{PhysicsReports}.
Proofs of existence and nonexistence
of breathers, as spatially localised and time-periodically varying solutions,   were
provided in \cite{MacKay}-\cite{Dirk} using different techniques. The exponential stability of breathers
was proven in \cite{Bambusi}.
Analytical and numerical methods have been
developed to continue breather solutions in conservative and dissipative systems
starting from the anti-integrable limit \cite{Sepulchere}-\cite{Martinez2}.

During recent years the
existence of ILMs (breathers) has been verified in a number
of experiments in various contexts including micro-mechanical
cantilever arrays \cite{cantilever},
arrays of coupled Josephson junctions \cite{josephson}, antiferromagnetic chains \cite{antiferro}, coupled optical wave guides \cite{optical},
Bose-Einstein condensates in optical lattices \cite{BEC}, in coupled torsion pendula \cite{Jesus},
electrical transmission lines \cite{electrical}, and granular crystals \cite{crystals}.
Regarding their creation mechanism in conservative systems, modulational instability (MI) provides the route to
the formation of breathers originating from an initially spatially homogeneous state imposed to (weak) perturbations.
To be precise, the MI of band edge plane waves
triggers an inherent instability leading to the formation of a spatially  localised state
\cite{Remoissenet}.

Lattice discreteness and nonlinearity are the prerequisites for
the ability of breathers to store energy without dispersing it to the environs.
Discreteness serves for the existence of an upper boundary to the linear phonon spectrum making
it possible that not only the fundamental
breather frequency but also all of its higher harmonics lie outside of the phonon spectrum. Nonlinearity
allows to tune the frequency
of (anharmonic) oscillators in dependence of their amplitude (or action) such that resonances between harmonics
of the breather frequency, $\omega_b$, and the
frequencies, $\omega_{ph}$, of linear oscillations, i.e. phonons, can be avoided. That is, the non-resonance conditions
\begin{equation}
 m \omega_b \ne \omega_{ph},\,\,\,\forall m \in {\mathbb{Z}}
\end{equation}
need to be satisfied.

In this work we prove the existence of spatially localised time-periodic
solutions
for which the small amplitude
oscillations at the lattice sites are
characterised by a uniform
time-dependent functional form. The paper is organised as follows: In the next section we
introduce general nonlinear Klein-Gordon lattice system and discuss the features of 
time-reversibility. Subsequently the theorems together with their 
proofs of the existence of time-periodic 
and spatially localised solutions in KG systems with soft and hard on-site potentials are 
presented. It follows a section on the 
numerical analysis of the solution behaviour of lattice states 
with frequencies close to the edges of the linear phonon band. Finally we summarise our results.

\section{Nonlinear Klein-Gordon lattices}

We study the dynamics of general one-dimensional infinite nonlinear Klein-Gordon (KG) lattice systems
given by the following system of equations
\begin{eqnarray}
\frac{d^2q_n}{dt^2}&=&-U^{\prime}(q_n)
+\kappa (q_{n+1}-2q_n+q_{n-1}),\,\,\,n \in {\mathbb{Z}}
\label{eq:start}
\end{eqnarray}
where the prime $^{\prime}$ stands for the derivative with respect to $q_n$, the latter being
the coordinate of the oscillator at site $n$ evolving in an anharmonic on-site potential
$U(q_n)$.
Each oscillator is coupled to its next neighbours and the strength of the interaction
is regulated by the value of the parameter $\kappa$.
This system has a Hamiltonian structure related to the energy
\begin{equation}
 H=\sum_{n \in {\mathbb{Z}}}\left(\frac{1}{2}p_n^2+U(q_n)+
 \frac{\kappa}{2}(q_{n}-q_{n-1})^2\right),
\end{equation}
and it is
time-reversible with respect to the involution $p\mapsto -p$.

The on-site potential $U$ is analytic  and is assumed to have the following properties:
\begin{equation}
U(0)=U^{\prime}(0)=0\,,\,\,\,U^{\prime \prime}(0)> 0.\label{eq:assumptions}
\end{equation}
Furthermore, we assume that $U(q)=U(-q)$.
In what follows we differentiate between soft on-site potentials and hard on-site
potentials. For the former (latter)
the oscillation frequency of an oscillator
decreases (increases) with increasing oscillation amplitude.
A soft potential possesses (at least) two inflection points
at $-q_{i}<0$ and $q_{i}>0$  and it holds that
\begin{eqnarray}
U^{\prime}(-q_{i}<q<0)&<&0,\,\,\,  U^{\prime}(0<q<q_{i})>0\label{eq:U1prime}\\
U^{\prime \prime}(\pm q_{i})&=&0,\,\,\,  U^{\prime \prime}(-q_{i}<q<q_{i})>0.\label{eq:U2prime}
\end{eqnarray}
We remark that $U(q)$ can have more than two inflection points (an example is the periodic potential $U(q)=-\cos(q)$).
In the frame of the current study we are only interested in motion
between the inflection points adjacent to the minimum of $U(q)$ at $q=0$.

Hard on-site potentials are, in addition to the assumptions in (\ref{eq:assumptions}),
characterised in their entire range of definition  by
\begin{equation}
U^{\prime}(q<0)<0,\,\,\,U^{\prime}(q>0)>0,\,\,\,U^{\prime \prime}(q)> 0.\label{eq:assumptionhard}
\end{equation}

The linearised system admits plane wave (phonon) solutions with frequencies
\begin{equation}
  \omega_{ph}^2(k)= \omega_{0}^2+4\kappa\sin^2\left(\frac{k}{2}\right),\,\,\,k \in [-\pi,\pi]
  \label{eq:linear}
 \end{equation}
and $\omega_0^2=U^{\prime \prime}(0)$.

We study solutions to the system (\ref{eq:start}) that are
 time-periodic satisfying
 \begin{equation}
  q_n(t+T_b)=q_n(t)
 \end{equation}
with period $T_b=2\pi/\omega_b$. Moreover, these periodic solutions are supposed to be
localised about a
single site, say $n=0$, exhibiting the reversion symmetry
$q_{n}(t)=q_{-n}(t)$.

In compact notation the canonical equations
can be expressed as
\begin{equation}
 \dot{X}=F(X)\,,\label{eq:system}
\end{equation}
where $X=(X_n)_{n \in {\mathbb{Z}}}$ is the infinite sequence of pairs
$X_n=(p_n,q_n)$ of canonically conjugate momentum and coordinate variables and the vector field
$F=(F_n)_{n \in {\mathbb{Z}}}$ is determined by
\begin{equation}
 F_n=(-\partial H/\partial q_n,\partial H/\partial p_n). \label{eq:reversible}
\end{equation}
In particular, due to the finite radius of the  interaction between the oscillators
(nearest-neighbour coupling)
the r.h.s. $F$ of
the system (\ref{eq:system})  is of the form
$F_n=F_n(X_{n-1},X_n,X_{n+1})$.

Introducing the involution
\begin{equation}
\rho:\,\,\,\left( \begin{array}{c} p_n \\ q_n \end{array} \right)\,
\mapsto \left( \begin{array}{c} -p_n \\ q_n \end{array} \right)\,,
\end{equation}
the time-reversibility of the system (\ref{eq:system}) is reflected in
\begin{equation}
 \rho F(X)=-F(\rho(X)).
\end{equation}

The forthcoming results in relation to
the existence of localised periodic solutions to
(\ref{eq:start}) are supported by the following statement.

\vspace*{0.5cm}
\noindent {\bf Lemma:} {\it Let $X(t)=(X_n(t))_{n \in {\mathbb{Z}}}$
 be a solution to the time-reversible system (\ref{eq:system}) with the reversible vector field
 (\ref{eq:reversible}) and
 with initial conditions $X(0)=(X_n(0))_{n \in {\mathbb{Z}}}$
 such that $p_n(0) \ne 0$ and $q_n(0)= 0$
 for ${n \in {\mathbb{Z}}}$
 and there is
 some $T > 0$ such that $\rho X(0) = X(T)$.
 Then on the interval $[0,T]$ one has
 $\rho X(t) = X(T-t)$, i.e. $X(t)$ is reversible on $[0,T]$.}

\vspace*{0.5cm}

\noindent{\bf Proof:}
The solution to $\dot{X}=F(X)$
with initial conditions $X(0)$ for $t\in [0,T]$ is expressed as
\begin{equation}
 X(t)=X(0)+\int_{0}^{t} d\tau F(X(\tau)).
\end{equation}
Then, using the reversibility properties $\rho X(0) = X(T)$ and $\rho F(X) = -F(\rho X)$, one gets
\begin{eqnarray}
 \rho X(t)&=&\rho X(0)+\rho\int_{0}^{t}d\tau F(X(\tau))\nonumber\\
 &=&X(T)+\int_{0}^{t}d\tau \rho F(X(\tau))\nonumber\\
 &=&X(T)-\int_{0}^{t}d\tau F(\rho X(\tau))\nonumber\\
 &=&X(T-t)\nonumber
\end{eqnarray}
and the proof is complete.

\hspace{16.5cm} $\square$

The next section deals with the existence of localised solutions.

\vspace{0.5cm}

\section{Existence of localised solutions}

In the following we prove the existence of spatially localised and
time-periodic solutions
for the system\,(\ref{eq:start}) where the oscillators perform collective motion
of the form $q_n(t)=\phi_n f(t)$ with the sequence $(\phi_n)_{n\in {\mathbb{Z}}}$
determining the amplitudes
of periodic oscillations, the latter being represented by a periodic function
$f(t)=f(t+T_b)$ of period $T=2\pi/\omega_b$ \cite{GTS}.   Localisation requires
$\lim_{|n|\rightarrow + \infty}\| \phi_n-\phi_{n-1}\|_{l^{\infty}}\,=0$.
These solutions, like  normal modes, are understood as a
{\it vibration in unison} of the system, i.e.
all units of the system perform synchronous oscillations
so that all the oscillators pass through their extreme values simultaneously
\cite{Rosenberg},\cite{Vakakis}. The latter feature is also common to
discrete breather patterns arising in
nonlinear lattice systems, where the oscillators
perform in-phase and/or out-of-phase single-frequency periodic motion  with respect to a
reference oscillator \cite{Aubry1},\cite{Morgante}.
By comparison to the uniform time-dependence of NNMs, in general for discrete breathers
the functional form of the time-dependence changes from site to site.
Nevertheless, for  small amplitude localised solutions 
in infinite lattices
with frequencies of oscillations, $\omega_b$, near the end points
of the spectral phonon band (i.e.
close to the lowest (highest) normal mode frequency
in systems with soft (hard) on-site potentials, i.e.
$\omega_b \lessapprox \omega_0$ ($\omega_b \gtrapprox \sqrt{\omega_0^2+4\kappa}$))
solutions are well approximated in the form $q_n(t)=\phi_n f(t)$ \cite{GTS}.
In \cite{James1},\cite{James2}, using a center manifold technique,  it is proven
that small amplitude breather solutions in one-dimensional infinite nonlinear lattice
systems, possessing a band formed by the frequencies
of the linear spectrum, exist and are of the form $q_n(t)=b_n f(t)$ where $f(t)$ is a periodic harmonic
function in time
and the decaying amplitudes $b_n$ constitute an exponentially localised shape
as $|n|\rightarrow \infty$.
Such discrete breathers of small amplitude emerge from a bifurcation of a band
edge plane wave.
 In the limit of their amplitude approaching zero the degree of localisation
 (localisation length) goes to zero.
Furthermore, approximations of small amplitude breathers in KG lattice systems near
the end points of the spectral band
by solitons of  discrete nonlinear Schr\"odinger
equations can be obtained  by means of a
resonant normal form theorem \cite{Penati}.
We represent  here a proof  of
the existence of small amplitude time-periodic and spatially localised
solutions in infinite one-dimensional KG lattices by means of an alternative technique.

\vspace*{0.5cm}

Based on the work in \cite{James1},\cite{James2}, we make the {\bf Assumption}
that 
lattice states with frequencies
close to the phonon band edges are well approximated
by  $q_n(t)=\phi_{n} f(t)$ with $n \in {\mathbb{Z}}$,
and state the main results of the paper showing
that the oscillators perform in-phase (out-of-phase) periodic motions in 
systems with a soft (hard) on-site potential 
with decaying 
amplitudes $\phi_{|n|}> \phi_{|n+1|}$,\, \,\,$\pm n\in {\mathbb{N}}$  forming a localised pattern.

\vspace*{0.5cm}
\noindent {\bf Theorem 1:} {\it Consider system (\ref{eq:start}) 
with a soft on-site potential.
Suppose the solutions  
corresponding to lattice states with frequencies close to the lower edge 
of the linear phonon band  
 are of the form 
 \begin{equation}
 q_n(t)=\phi_{n} f(t),\,\,\,n\in {\mathbb{Z}},\,\,\,\phi_n \in {\mathbb{R}},\label{eq:ansatz}
\end{equation}
fulfilling
$|q_n(t)|\le \xi<q_i$ for $t \in {\mathbb{R}}$ where $\xi$ and $q_i$ denote the 
maximal amplitude of motion
and
the (positive) inflection point of the soft potential respectively.
Provided the inequality
\begin{equation}
U^{\prime \prime}(\xi)>\kappa A,\label{eq:rel1}
\end{equation}
where
\begin{equation}
A=\max_{n\in {\mathbb{Z}}^+}
\left\{\frac{(\phi_{n+1}-2\phi_n+\phi_{n-1})-(\phi_{n+2}-2\phi_{n+1}+\phi_n)}
{\phi_n-\phi_{n+1}}\right\}\label{eq:A}\\
\end{equation}
is satisfied then 
there exist localised periodic solutions
$(q_n(t+T_b),\dot{q}_n(t+T_b))=(q_n(t),\dot{q}_n(t))$,
so that
the oscillators perform in-phase motion, i.e.
$\sign (q_n)=\sign (q_{n+1})$, given by
with $0<\phi_{n+1}<\phi_n$ for $n\ge 0$ and
$\phi_{n}>\phi_{n-1}>0$ for $n< 0$.
$f$ is a
periodic smooth function of time $t$ with period $T_b=2\pi/\omega_b$ and frequency
$\omega_b$ satisfying
\begin{equation}
\omega_b < \omega_0. \label{eq:ineq1}
\end{equation}
For time-reversible solutions $f(t)$ is determined by $f(t)=\cos(\omega_b t)$.
}

\vspace*{0.5cm}
\noindent {\bf Proof:} W.l.o.g. the initial conditions are chosen such that
$q_{n}(0)=0$,  and
$\dot{q}_{n}(0) \ne 0$,  say $\dot{q}_{n}(0) > 0$,  $\forall n$,
and $\dot{q}_{n+1}(0)>\dot{q}_{n}(0)>0$ for $n < 0$, and
$\dot{q}_{n}(0)>\dot{q}_{n+1}(0)>0$ for $n \ge 0$.
Then due to continuity there exists some
$t_*>0$ so that during the interval $(0,t_*]$ the
following  order relations are satisfied
\begin{eqnarray}
 0 &<& q_{n+1}(t) < q_{n}(t)\,,\,\,\,n \ge 0,\label{eq:order1}\\
 0 &<& q_{n}(t) < q_{n+1}(t)\,,\,\,\,n < 0.\label{eq:order2}
\end{eqnarray}

We define the difference variable between the coordinates and velocities
respectively at sites $n$ and $n+1$ as follows
\begin{equation}
\Delta q_{n}(t)= \alpha [q_{n}(t)-  q_{n+1}(t)],\,\,\,\Delta \dot{q}_{n}(t)=
 \alpha[\dot{q}_{n}(t)-  \dot{q}_{n+1}(t)],\label{eq:def}
\end{equation}
where $\alpha =1$ ($\alpha =-1$) for $n\ge 0$ ($n< 0$).
Thus by definition  $\Delta q_n(t) \ge 0$ on $[0,t_*]$.

Due to reversion symmetry $q_n=q_{-n}$ it suffices to consider
$n\ge 0$, viz. $\alpha=1$.
The time evolution of the difference variables  $\Delta q_n (t)$ is
determined by the following equation
\begin{equation}
 \frac{d^2 \Delta q_n}{dt^2}= -\left[U^{\prime}(q_{n})-U^{\prime}(q_{n+1})\right]
 +{\kappa}(\Delta q_{n+1}- 2\Delta q_{n}+\Delta q_{n-1})\label{eq:deltap1}.
\end{equation}

Utilising that for $q_k\ge q_l$ and $|q_n|_{n \in \mathbb{Z}}\le \xi<q_i$ it holds that
\begin{equation}
 U^{\prime \prime}(0)(q_k-q_l)\ge U^{\prime}(q_k)-U^{\prime}(q_l)\ge
 U^{\prime \prime}(\xi) (q_k-q_l)> 0
\end{equation}
 enables us
to bound the evolution of the difference variable $\Delta q$  on $[0,t_*]$
from above as
\begin{equation}
 \frac{d^2 \Delta q_n}{dt^2} \le  -U^{\prime \prime}(\xi) \Delta q_n
 +{\kappa}(\Delta q_{n+1}- 2\Delta q_{n}+\Delta q_{n-1})\,\,,
 \label{eq:deltapa}
\end{equation}
and from below as
\begin{equation}
 \frac{d^2 \Delta q_n}{dt^2} \ge  -\omega^2_0 \Delta q_n
 +{\kappa}(\Delta q_{n+1}- 2\Delta q_{n}+\Delta q_{n-1}).\label{eq:deltapb}
\end{equation}

Substituting $q_n(t)=\phi_{n} f(t)$
into Eqs.\,(\ref{eq:deltapa}) and (\ref{eq:deltapb}) gives
\begin{equation}
\frac{d^2 f}{dt^2}\le
-\left(U^{\prime \prime}(\xi)-\kappa
\left[\frac{(\phi_{n+1}-2\phi_n+\phi_{n-1})-(\phi_{n+2}-2\phi_{n+1}+\phi_n)}
{\phi_n-\phi_{n+1}}
\right]\right)f\le -\left(U^{\prime \prime}(\xi)-\kappa A\right)f,
\label{eq:osc1}
\end{equation}
and
\begin{equation}
\frac{d^2 f}{dt^2}\ge
-\left(\omega^2_0-\kappa \left[\frac{(\phi_{n+1}-2\phi_n+\phi_{n-1})-(\phi_{n+2}-2\phi_{n+1}+\phi_n)}
{\phi_n-\phi_{n+1}}
\right]\right)f\ge -\left(\omega_0^2-\kappa B\right)f.\label{eq:osc2}
\end{equation}
 with $A$  given in (\ref{eq:A}) and $B$ is determined by
\begin{equation}
B=\min_{n\in {\mathbb{Z}}^+}\left(\frac{(\phi_{n+1}-2\phi_n+\phi_{n-1})-(\phi_{n+2}-2\phi_{n+1}+\phi_n)}
{\phi_n-\phi_{n+1}}\right)\label{eq:B}.
\end{equation}
In  compliance with the order relations (\ref{eq:order1}) the amplitudes
$\phi_n$ form a localised sequence $0<\phi_{n+1}<\phi_n$ for $n\ge 0$.

Therefore, by the comparison principle for differential equations, for given initial conditions,
$f(t)$ and $df(t)/dt=\dot{f}(t)$ are bounded from above and below
 by the solution of
\begin{equation}
\frac{d^2 G}{dt^2}=-\tilde{\Omega}_s^2  G
\label{eq:boundabove}
\end{equation}
and
\begin{equation}
\frac{d^2 g}{dt^2}=-\tilde{\omega}_s^2 g,
\label{eq:boundbelow}
\end{equation}
respectively, provided $G(t)\ge 0$ and $g(t)\ge 0$ and we introduced
\begin{eqnarray}
\tilde{\Omega}_s^2&=&U^{\prime \prime}(\xi)-\kappa A,\\
\tilde{\omega}_s^2&=&\omega^2_0-\kappa B\label{eq:freqs}.
\end{eqnarray}

By the hypothesis (\ref{eq:rel1}) it is assured that
$\tilde{\Omega}_s \in {\mathbb{R}}$.
Furthermore, since
$\omega_0>U^{\prime \prime}(\xi)$   it follows that
$\tilde{\omega}_s\in {\mathbb{R}}$ and so there exist
oscillatory solutions to (\ref{eq:boundabove}) and
(\ref{eq:boundbelow}). 

The solution to Eq.\,(\ref{eq:boundabove}) and (\ref{eq:boundbelow})
with initial conditions $(f(0)=0, \dot{f}(0)>0)$ (complying with
$q_n(0)=0$ and $\dot{q}_n(0) \neq 0$
for all $n$) is given by
\begin{eqnarray}
  G(t)&=& \frac{\dot{f}(0)}{\tilde{\Omega}_s}\sin(\tilde{\Omega}_s t),\\ \label{eq:qabove1}
  \dot{G}(t)&=& \dot{f}(0)\cos(\tilde{\Omega}_s t)
  \label{eq:pabove}
\end{eqnarray}
and
\begin{eqnarray}
  g(t)&=& \frac{\dot{f}(0)}{\tilde{\omega}_s}\sin(\tilde{\omega}_s t),\\ \label{eq:qbelow1}
  \dot{g}(t)&=& \dot{f}(0)\cos({\tilde{\omega}_s} t)
  \label{eq:pbelow}
\end{eqnarray}
respectively, where
$G(t)\ge 0$ for $0 \le t \le  \pi/\tilde{\Omega}_s$ and $g(t)\ge 0$
for $0 \le t \le  \pi/\tilde{\omega}_s$.

Notice that $d^2 f(t)/dt^2< 0$ on $(0,t_*)$, that is,
the acceleration stays negative.
Due to the relations $f>0$ on $(0,t_*)$ in conjunction with the lower bound
   $g(t) \le f(t)$  the order relations
as given in (\ref{eq:order1}),(\ref{eq:order2})  are at
least maintained on the interval
$(0,\pi/\tilde{\omega}_s)$. Moreover,
$f(t)$
is bound to grow monotonically
at least during the interval
$(0,\pi/(2\tilde{\omega}_s)]$
and attains a least maximal value $\dot{f}(0)/\tilde{\omega}_s$.
Furthermore, $f(t)$ cannot return to zero before
$t=\pi/\tilde{\omega}_s$.

On the other hand, from the upper bound $ f(t) \le G(t)$  one infers that $f(t)$
can attain an absolute maximal value
$\dot{f}(0)/\tilde{\Omega}_s$ but not before $t=\pi/(2\tilde{\Omega}_s)$ and
$f(t)$ is bound
to return to zero not later than $t=\pi/\tilde{\Omega}_s$.

Therefore,  there must be a $t_* \in [\pi/{\tilde{\omega}_s},\pi/{\tilde{\Omega}_s}]$
such that $f(t_{*})=f(0)=0$ and
$\dot{f}(t_{*})=-\dot{f}(0)$.

Hence one has $q_n(t_*)=q_n(0)$ and $\dot{q}_n(t_*)=-\dot{q}_n(0)$,
 and by the Lemma above their time evolution
is given in terms of  reversible
solutions obeying
\begin{eqnarray}\
q_n(t_{*}/2+t)&=&q_n(t_{*}/2-t),\\
\dot{q}_n(t_{*}/2+t)&=&-\dot{q}_n(t_{*}/2-t)),
\end{eqnarray}
with $0\le t \le t_{*}/2$ and at the turning point, $t=t_{*}/2$,
the coordinates
$q_{n}$
assume simultaneously their
respective maxima
while  the $\dot{q}_{n}$ are all zero. Conclusively, on the interval $[0,t_*]$
the  oscillators evolve through half a  cycle of periodic
in-phase motion, i.e. $\sign (q_n)=\sign (q_{n+1})$ and
$\sign(\dot{q}_n)=\sign(\dot{q}_{n+1})$.
At  $t=t_*$ one has $q_n(t_*)=q_n(0)=0$ and $\dot{q}_n(t)=-\dot{q}_n(0)$ so that
by symmetry during the interval $[t_*,2t_*]$ the oscillators repeat their
behaviour exhibited during the initial interval
$[0,t_*]$ but this time with
reversed sign of their coordinates and velocities and hence, at $t=2t_*$ one period of localised periodic oscillations
is completed.
Repeating the  process results in time-periodic and spatially localised motions
of the oscillators. 

By taking the values of the coordinates $q_{n}$
at the (first) turning point $t=t_*/2$ together with $\dot{q}_{n}=0$
as (new) initial conditions one generates  time-reversible solutions
possessing the symmetry $q_{n}(t)=q_{n}(-t)$
and $-\dot{q}_{n}(t)=\dot{q}_{n}(-t)$ so that
the  order relations
\begin{eqnarray}
 |q_{n+1}(t)|&\le& |q_n(t)|,\,\,\,n \ge 0,\label{eq:test3}\\
 |q_{n+1}(t)|&\ge& |q_n(t)|,\,\,\,n<0 \label{eq:test4}
\end{eqnarray}
are true for $t\in {\mathbb{R}}$. As $\lim_{|n|  \rightarrow \infty} q_n =0$ 
the tails of
the breather solutions behave
linearly so that their time-dependence is characterised by a single-frequency
harmonic function. The latter for time-reversible solutions
 is determined by $f(t)=\cos(\omega_b t)$. 
 
 Furthermore, the amplitudes in the exponentially localised 
 tails of the breathers can be represented as 
 $\phi_{|n|}=K \lambda^{|n|}$ with $0<\lambda <1$, $\pm n \in {\mathbb{N}}$ and $K>0 \in {\mathbb{R}}$ and 
 when substituted in (\ref{eq:A}) one gets 
 \begin{equation}
\frac{(\phi_{n+1}-2\phi_n+\phi_{n-1})-(\phi_{n+2}-2\phi_{n+1}+\phi_n)}
{\phi_n-\phi_{n+1}}=K\left(\lambda+\frac{1}{\lambda}-2\right)>0\,\,\,{\rm{for}}\,\,\,0<\lambda <1.
 \end{equation}
Then concerning the frequencies of oscillations one finds
from (\ref{eq:freqs}) that
\begin{equation}
\omega_b \le \sqrt{\omega_0^2-\kappa\left[\frac{(\phi_{n+1}-2\phi_n+\phi_{n-1})-(\phi_{n+2}-2\phi_{n+1}+\phi_n)}
{\phi_n-\phi_{n+1}}
\right]}<\omega_0
\end{equation}
and thus the breather frequencies lie below the lower edge of the linear band.
 
Note that the inequalities in (\ref{eq:test3}) and (\ref{eq:test4}) are  strict except  at those moments in time,
$\tilde{t}_{k\in {\mathbb{Z}}}$, when the
 oscillators pass simultaneously through zero coordinate,
corresponding to the minimum position of the on-site potential, i.e.
$q_{n}(\tilde{t}_k)=0$ and the proof is complete.

\hspace{16.5cm} $\square$

\vspace*{1.cm}

\noindent {\bf Remark:}
The relations
\begin{eqnarray}
 & &\sqrt{U^{\prime \prime}(\xi)-\kappa
 \max_{n \in {\mathbb{Z}}^+}\left\{\frac{(\phi_{n+1}-2\phi_n+\phi_{n-1})-(\phi_{n+2}-2\phi_{n+1}+\phi_n)}
{\phi_n-\phi_{n+1}}\right\}}
 \le \omega_b\nonumber\\
 &\le&
 \sqrt{\omega_0^2-\kappa \min_{n \in {\mathbb{Z}}^+}\left\{\frac{(\phi_{n+1}-2\phi_n+\phi_{n-1})-(\phi_{n+2}-2\phi_{n+1}+\phi_n)}
{\phi_n-\phi_{n+1}}\right\}}\label{eq:rel5}
\end{eqnarray}
entail that
\begin{equation}
 \lim_{\xi \rightarrow 0}\,\lim_{(\phi_{n})_{n \in \mathbb{Z}}^+ \rightarrow C}\,\omega_b=\omega_0\,,\,\,\,
 C \in {\mathbb{R}}.
\end{equation}
This is in compliance with the fact that breathers of 
(arbitrarily) small amplitude exists
in one-dimensional infinite anharmonic lattices near the (lower) edge of the linear band of phonons. 
In fact, small amplitude
breathers emerge
from band edge plane waves by means of a tangent bifurcation \cite{James}.
Conversely, in the limit $\phi_n \rightarrow C$ with $C \in {\mathbb{R}}$, i.e. when
the localisation strength approaches zero, the breather is
transformed into a spatially extended solution, viz. a plane wave.

Moreover, the condition (\ref{eq:rel5}) involving the difference between the
(local) curvatures
$(\phi_{n+1}-2\phi_n+\phi_{n-1})$ and
$(\phi_{n+2}-2\phi_{n+1}+\phi_{n})$
imposes, in dependence of the maximal
amplitude of motion,\,$\xi$, and the coupling strength $\kappa$,
a constraint  on the (local) degree of localisation.

\vspace*{1.cm}

The next Theorem establishes the existence of out-of-phase localised solutions
for motions in hard on-site potentials.

\vspace*{0.5cm}
\noindent {\bf Theorem 2:}
{\it Consider system (\ref{eq:start}) with a hard on-site potential. Suppose  
the solutions 
corresponding to lattice states with frequencies close to the upper edge 
of the linear phonon band  
are of the form 
 \begin{equation}
 q_n(t)=(-1)^n\phi_{n} f(t),\,\,\,n\in {\mathbb{Z}},\,\,\,\phi_n \in {\mathbb{R}},\label{eq:ansatz1}
\end{equation}
fulfilling $|q_n(t)|\le \eta$ for
$t \in {\mathbb{R}}$ where $\eta$ denotes the  maximal amplitude of motion.
There exist localised periodic solutions $(q_n(t+T_b),\dot{q}_n(t+T_b))=(q_n(t),\dot{q}_n(t))$,
so that
the oscillators perform out-of-phase motion, i.e.
$\sign (q_n)=-\sign (q_{n+1})$ given by (\ref{eq:ansatz1}) 
and $0<\phi_{n+1}<\phi_n$ for $n\ge 0$ and
$\phi_{n}>\phi_{n-1}>0$ for $n< 0$.
$f$ is a
periodic smooth function of time $t$ with period $T_b=2\pi/\omega_b$ and frequency
$\omega_b$ satisfying
\begin{equation}
\omega_b > \sqrt{\omega_0^2+4\kappa}. \label{eq:ineq2}
\end{equation}}

\vspace*{0.5cm}
\noindent {\bf Proof:} W.l.o.g. the initial conditions satisfy
\begin{eqnarray}
 q_{n}(0)&=&0,\,\,\, 0<(-1)^{n+1}\dot{q}_{n+1}(0)<(-1)^n\dot{q}_{n}(0)
 \,,\,\,\,\,n\ge 0,\\
q_{n}(0)&=&0,\,\,\, 0<(-1)^{n}\dot{q}_{n}(0)<(-1)^{n+1}\dot{q}_{n+1}(0)\,,
\,\,\,\,n< 0.
 \end{eqnarray}
Then due to continuity there exists some
$t_*>0$ so that during the interval $(0,t_*]$ the oscillators  perform
out-of phase motion and the
following  order relations are satisfied
\begin{eqnarray}
0 &<& (-1)^{n+1} q_{n+1}(t) < (-1)^n q_{n}(t)\,,\,\,\,\,n\ge 0,\\\label{eq:order3}
0 &<& (-1)^{n} q_{n}(t) < (-1)^{n+1} q_{n+1}(t)\,,\,\,\,\,n< 0.\label{eq:order4}
\end{eqnarray}

We proceed as in the previous case for soft on-site potentials by introducing
the difference variable between
coordinates as in (\ref{eq:def}). The time
evolution of the difference variable is determined by an
equation similar to
(\ref{eq:deltap1}) and using that for $q_k>q_l$
it holds that
\begin{equation}
 U^{\prime \prime}(\xi)(q_k-q_l)\ge U^{\prime}(q_k)-U^{\prime}(q_l)\ge
 U^{\prime \prime}(0) (q_k-q_l)> 0
\end{equation}
we derive on $[0,t_*]$ bounds for evolution of the difference variable $\Delta q_{n}(t)$ an
upper bound and lower bound as
 \begin{equation}
\frac{d^2 \Delta q_n}{dt^2}\le -\omega^2_0 \Delta q_n
 +{\kappa}(\Delta q_{n+1}- 2\Delta q_{n}+\Delta q_{n-1})\label{eq:hardb},
\end{equation}
 and
\begin{equation}
 \frac{d^2 \Delta q_n}{dt^2}\ge -U^{\prime \prime}(\eta) \Delta q_n
 +{\kappa}(\Delta q_{n+1}- 2\Delta q_{n}+\Delta q_{n-1})\label{eq:harda}
\end{equation}
respectively and due to reversion symmetry $q_n=q_{-n}$
the further analysis can be restricted to
$n\ge 0$.

Substituting $q_n(t)=(-1)^n \phi_n f(t)$ into Eqs.\,(\ref{eq:hardb}) and (\ref{eq:harda})
yields
\begin{equation}
\frac{d^2 f}{dt^2}\le
-\left(\omega^2_0+\kappa
\left[\frac{\phi_{n+1}+\phi_{n+2}}{\phi_{n}+\phi_{n+1}}+
\frac{\phi_{n-1}+\phi_{n}}{\phi_{n}+\phi_{n+1}}+2
\right]\right)f\le -\left(\omega^2_0+\kappa C\right)f,
\label{eq:osc3}
\end{equation}
and
\begin{equation}
\frac{d^2 f}{dt^2}\ge
-\left(U^{\prime \prime}(\eta)+\kappa \left[\frac{
\phi_{n+1}+\phi_{n+2}}{\phi_{n}+\phi_{n+1}}+
\frac{\phi_{n-1}+\phi_{n}}{\phi_{n}+\phi_{n+1}}+2
\right]\right)f\ge -\left(U^{\prime \prime}(\eta)+\kappa D\right)f,\label{eq:osc4}
\end{equation}
with $C=\min_{n \in {\mathbb{Z}}^+}\left\{\frac{\phi_{n+1}+\phi_{n+2}}{\phi_{n}+\phi_{n+1}}+
\frac{\phi_{n-1}+\phi_{n}}{\phi_{n}+\phi_{n+1}}+2
\right\}$ and $D=\max_{n \in {\mathbb{Z}}^+}\left\{\frac{\phi_{n+1}+\phi_{n+2}}{\phi_{n}+\phi_{n+1}}+
\frac{\phi_{n-1}+\phi_{n}}{\phi_{n}+\phi_{n+1}}+2
\right\}$.

Thus, by the comparison principle the solutions $f(t)$ are sandwiched
as $h(t)\le f(t)\le H(t)$
where the upper and lower bound are given by
\begin{eqnarray}
  H(t)&=& \frac{\dot{f}(0)}{\tilde{\Omega}_h}\sin(\tilde{\Omega}_h t),\\ \label{eq:qabove}
  \dot{H}(t)&=& \dot{f}(0)\cos(\tilde{\Omega}_h t),
  \label{eq:paboveh}
\end{eqnarray}
and
\begin{eqnarray}
  h(t)&=& \frac{\dot{f}(0)}{\tilde{\omega}_h}\sin(\tilde{\omega}_h t),\\ \label{eq:qbelow}
  \dot{h}(t)&=& \dot{f}(0)\cos({\tilde{\omega}_h} t),
  \label{eq:pbelowh}
\end{eqnarray}
respectively, and
$H(t)\ge 0$ for $0 \le t \le  \pi/\tilde{\Omega}_h$ and $h(t)\ge 0$
for $0 \le t \le  \pi/\tilde{\omega}_h$, where
\begin{eqnarray}
 \tilde{\Omega}_h&=&U^{\prime \prime}(\eta)+\kappa D,\\
 \tilde{\omega}_h&=&\omega^2_0+\kappa C.
 \label{eq:freqh}
\end{eqnarray}

The remainder of the proof regarding the time periodicity of
the resulting localised solutions proceeds
in an analogous way as above for Theorem 1.

Conclusively,  spatially localised and time-periodic solutions
for out-of-phase motion in hard on-site potentials result which satisfy
\begin{eqnarray}
 |q_{n}(t)|\ge|q_{n-1}(t)|,\,\,\,n&\ge& 0,\label{eq:test}\\
 |q_{n+1}(t)|\ge |q_{n}(t)|,\,\,\,n&<&0\label{eq:test2}
\end{eqnarray}
for $t \in {\mathbb{R}}$
and $(q_n(t+T_b),\dot{q}_n(t+T_b))=(q_n(t),\dot{q}_n(t))$ for $n \in {\mathbb{Z}}$
with period $T_b=2\pi /\omega_b$.
The equality sign in (\ref{eq:test}) and (\ref{eq:test2}) is valid only at the instants of time, $\tilde{t}_{k\in {\mathbb{Z}}}$, when the
 oscillators pass simultaneously through zero coordinate,
corresponding to the minimum position of the on-site potential, i.e.
$q_{n}(\tilde{t}_k)=0$.
Using similar arguments as in the part of the proof of Theorem $1$ that concerns
the 
frequencies of oscillations, $\omega_b$, 
one infers from (\ref{eq:freqh}) that breather
frequencies lie above the upper
edge of the linear band fulfilling the relation
\begin{equation}
\omega_b>\sqrt{\omega_0^2+4\kappa}.
\end{equation}
 which concludes the proof.

\hspace{16.5cm} $\square$

\vspace*{1.0cm}

\section{Numerical analysis}

In this section, we verify, by means of a numerical analysis, the
validity of assumptions (\ref{eq:ansatz}) and (\ref{eq:ansatz1}). 
To this aim, we calculate breathers by using the numerical methods developed
in e.g. \cite{BreathersAC}. For examples of soft and hard on-site potentials 
we consider the sine-Gordon potential: $U(q_n)=1-\cos(q_n)$ and $\phi^4-$potential:
$U(q_n)=q_n^2/2+q_n^4/4$, respectively.





We quantitatively describe the validity of the above assumptions 
by monitoring the two following coefficients related to the relative norms of the exact breather profile ($\{q_n(0)\}$) and its approximation ($\{\phi_n\}$):



\begin{equation}\label{eq:norms}
    N_\infty=1-\frac{||\phi_n||_\infty}{||q_n||_\infty}
\qquad    
    N_2=1-\frac{||\phi_n||_2}{||q_n||_2}
\end{equation}

Consequently, the lower the values of $N_\infty$ and $N_2$, the better the approximation made in (\ref{eq:ansatz}) and (\ref{eq:ansatz1}).

\begin{figure}[!ht]
\begin{center}
\begin{tabular}{cc}
(a) & (b)\\
\includegraphics[width=.45\textwidth]{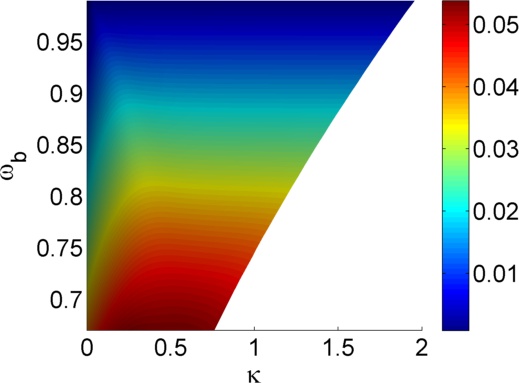} &
\includegraphics[width=.45\textwidth]{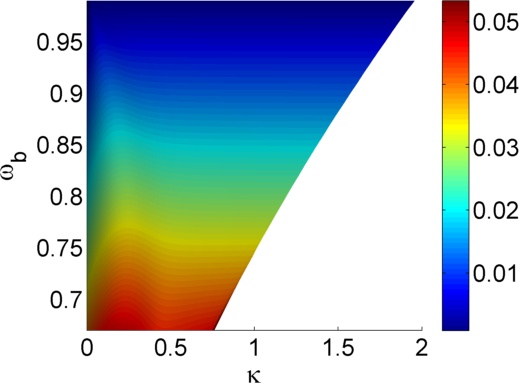}\\
(c) & (d) \\
\includegraphics[width=.45\textwidth]{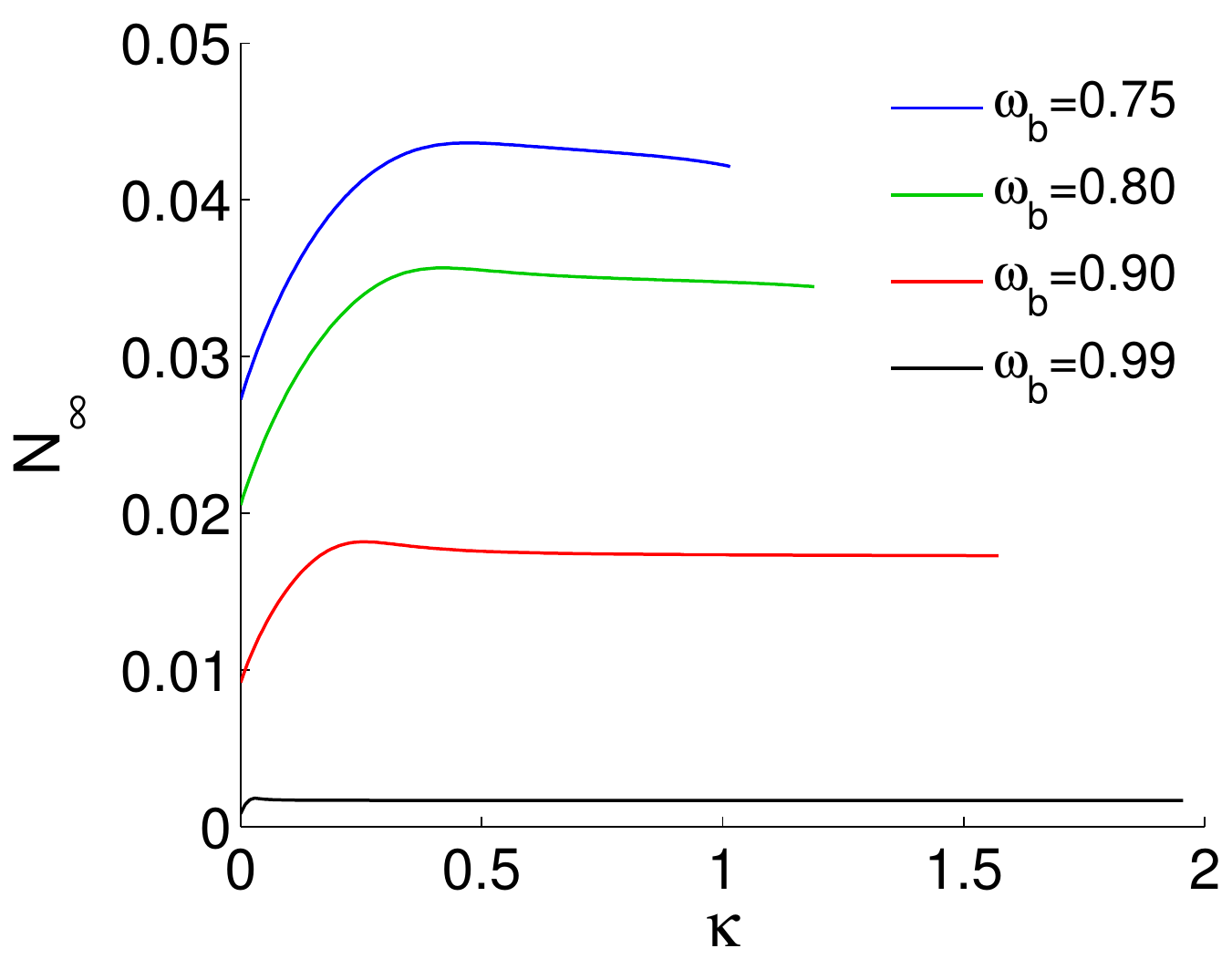} &
\includegraphics[width=.45\textwidth]{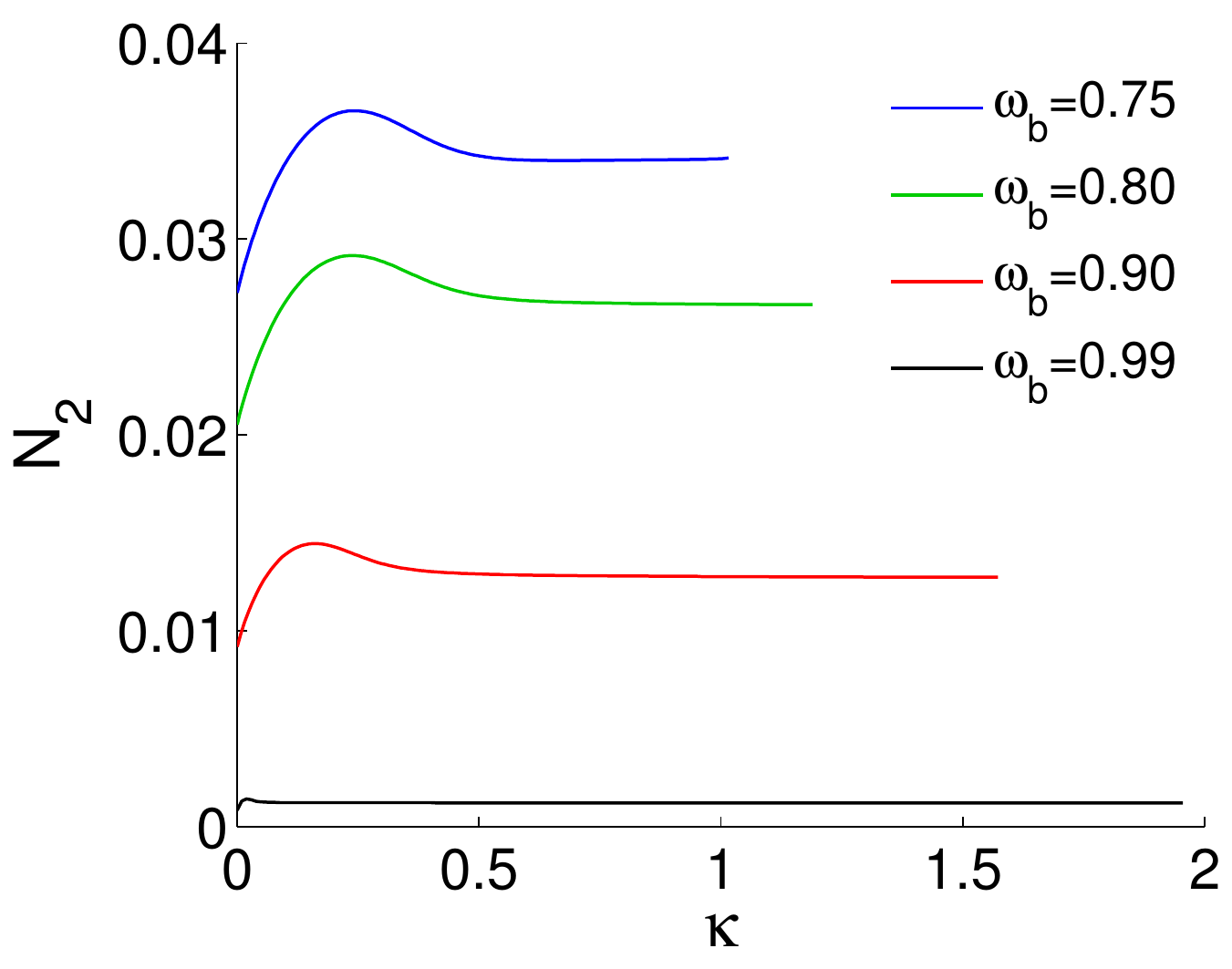}\\
\end{tabular}
\end{center}
\caption{The top panels show the dependence of the coefficients $N_\infty$ and $N_2$ with respect to the frequency $\omega_b$ and the coupling constant $\kappa$ for the sine-Gordon potential; notice that $\omega_0=1$. 
An alternative representation of the dependence of the coefficients with the coupling constant for four selected frequencies is depicted in the bottom panels.}%
\label{fig1}
\end{figure}

Figure \ref{fig1} shows the dependence of the validity coefficients with respect to the frequency $\omega_b$ and coupling constant $\kappa$ for the sine-Gordon potential. Notice that the maximum value of $\kappa$ is attained when $3\omega_b$ coincides with the upper edge of the phonon band. From this figure, we can observe that, for a given value of the frequency, 
the minima coefficients occurs at the anti-continuum limit (that is $\kappa=0$); 
in addition, the coefficients exhibit a relative maximum at a given value of the coupling that decreases with the frequency. Finally, for a given value of the coupling constant, the coefficients decrease when the frequency approaches the phonon band.

\begin{figure}[!ht]
\begin{center}
\begin{tabular}{cc}
(a) & (b)\\
\includegraphics[width=.45\textwidth]{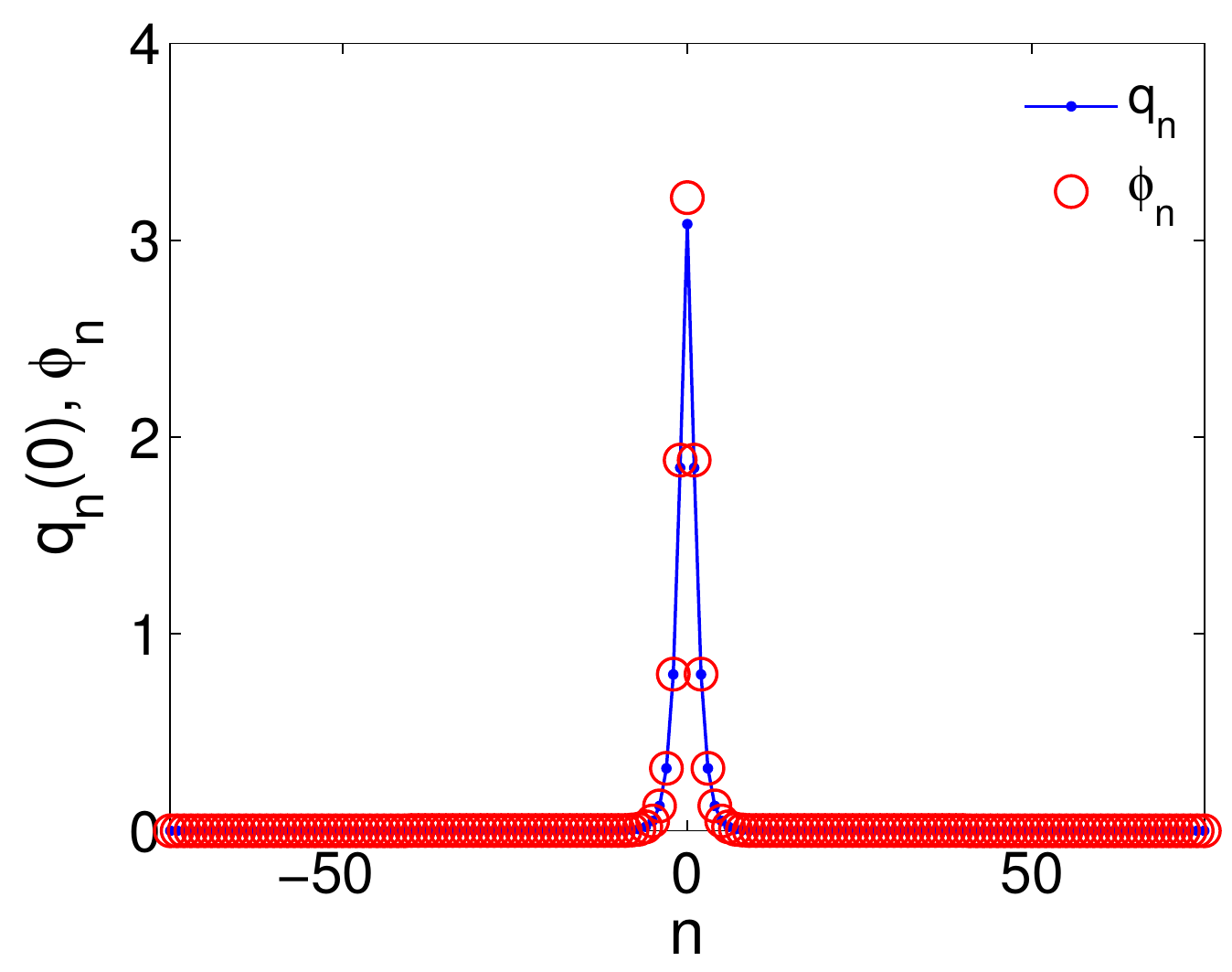} &
\includegraphics[width=.45\textwidth]{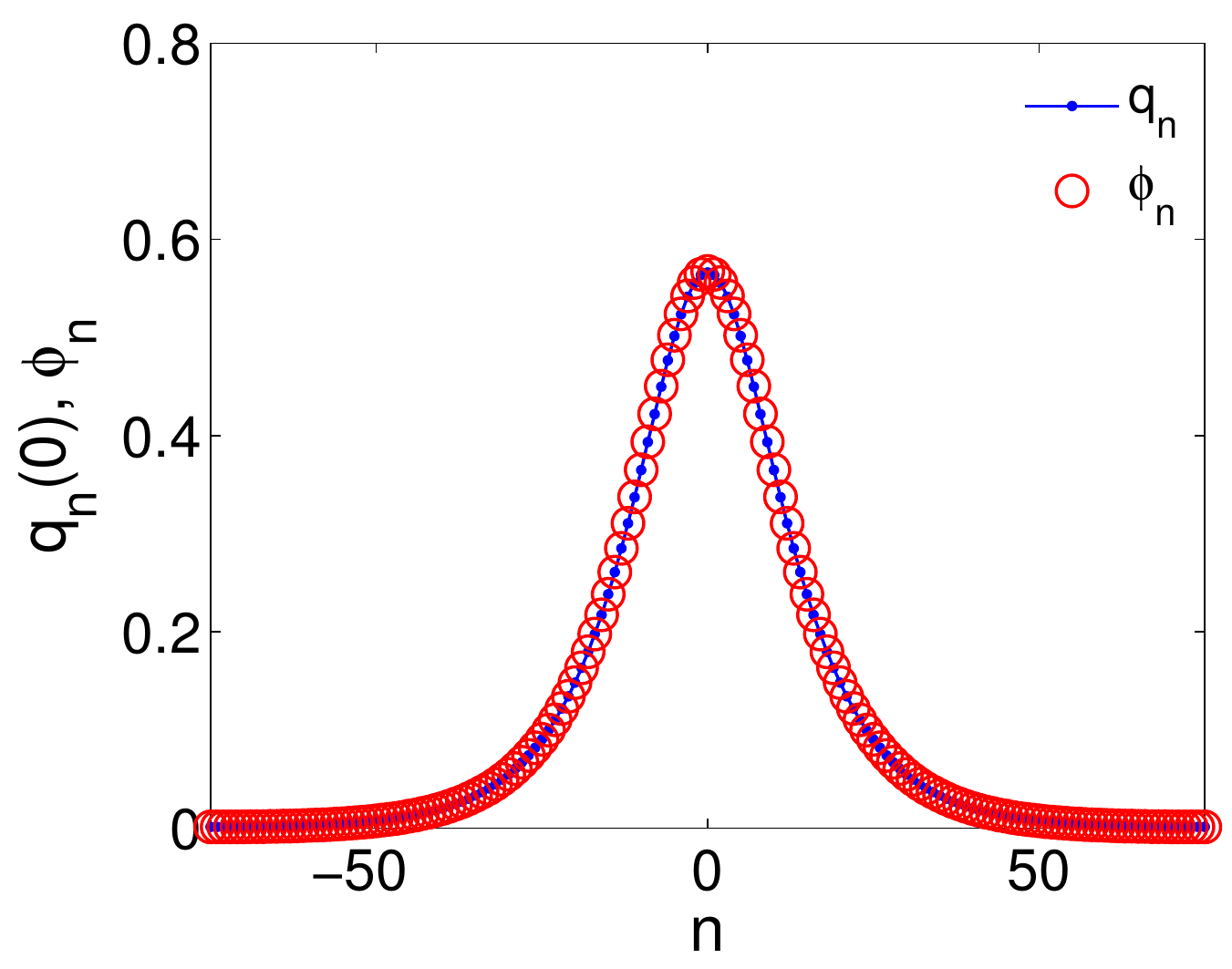}\\
\end{tabular}
\end{center}
\caption{Breather profiles for the sine-Gordon potential in two extreme cases. Panel (a) corresponds to a breather with $\omega_b=0.75$ and $\kappa=0.47$ whose validity coefficients are $N_\infty=0.0436$ and $N_2=0.0344$. Panel (b) shows a breather with $\omega_b=0.99$ and $\kappa=1.95$, characterized by $N_\infty=0.0017$ and $N_2=0.0012$. Blue line and dots correspond to the exact profile $\{q_n(0)\}$ and the approximated solution $\{\phi_n\}$ is depicted by red circles.}%
\label{fig2}
\end{figure}

A comparison of the breather profiles with the corresponding approximation 
is shown in Fig. \ref{fig2}, where two examples of 
two extreme cases (i.e. one for which the coefficients are maximal and another one for which the coefficients are minimal) are depicted. 
There is excellent agreement except at $n=0$.

\begin{figure}[!ht]
\begin{center}
\begin{tabular}{cc}
(a) & (b)\\
\includegraphics[width=.45\textwidth]{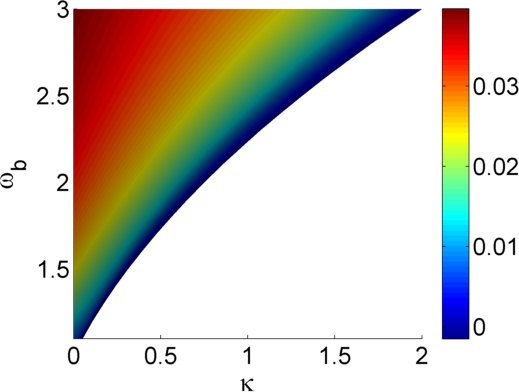} &
\includegraphics[width=.45\textwidth]{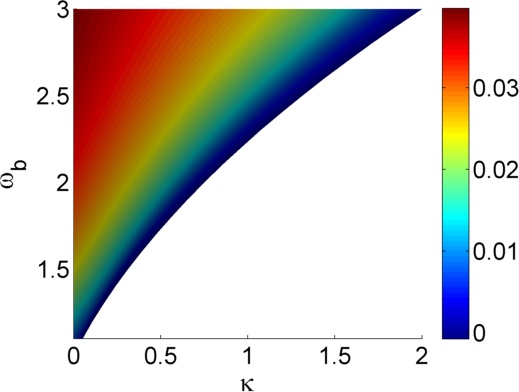}\\
(c) & (d) \\
\includegraphics[width=.45\textwidth]{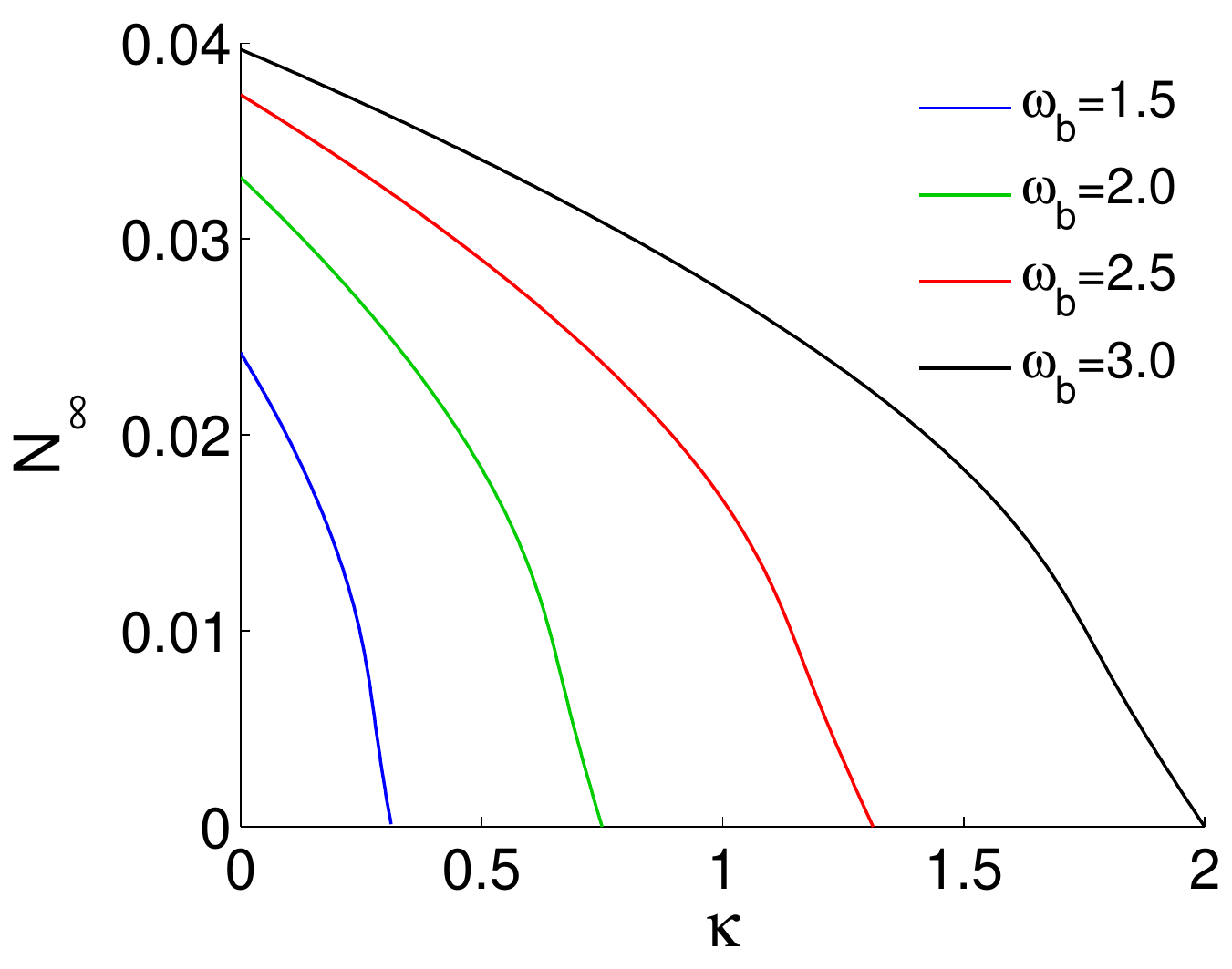} &
\includegraphics[width=.45\textwidth]{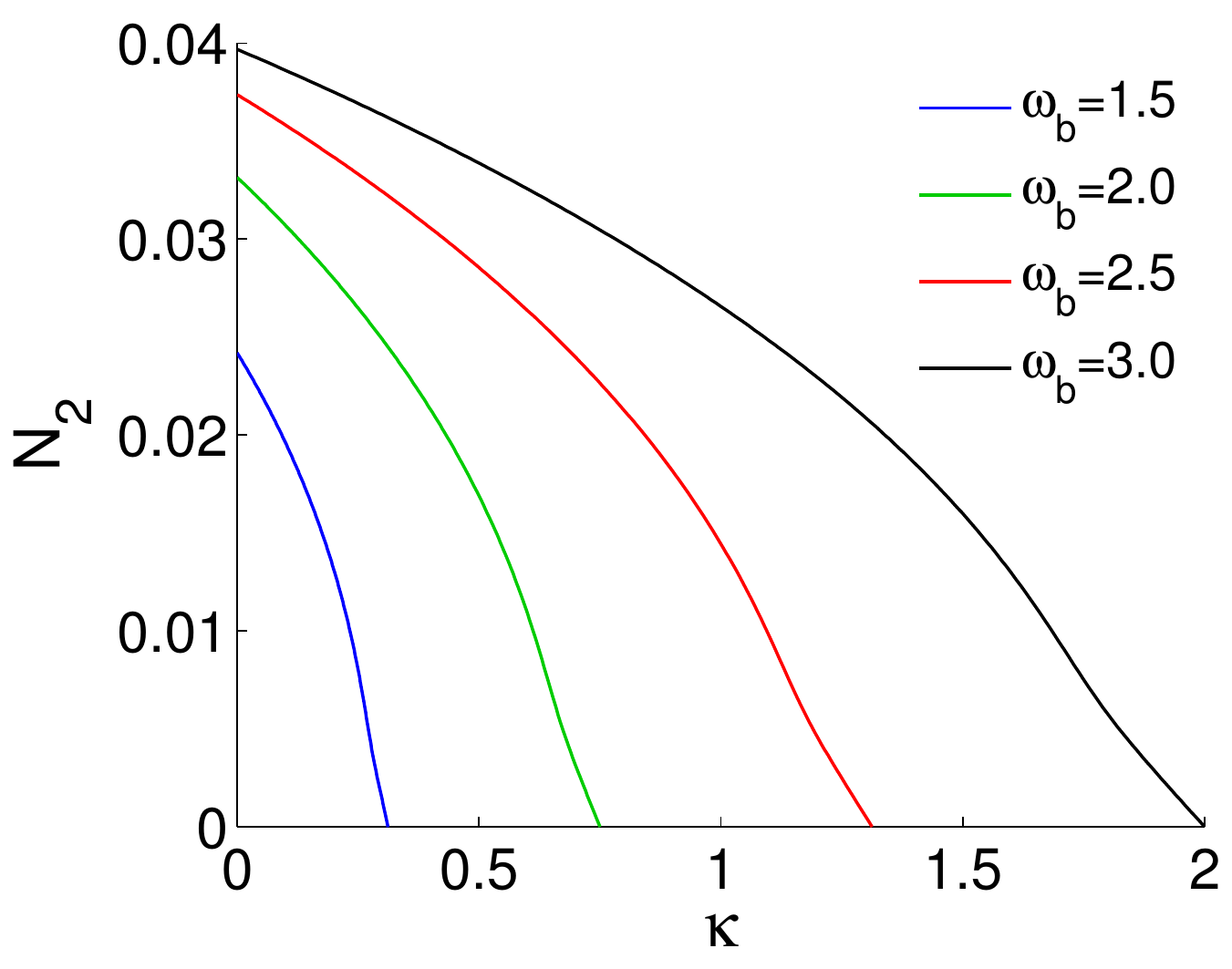}\\
\end{tabular}
\end{center}
\caption{Same as Fig. \ref{fig1} but for the hard $\phi^4$ potential.}%
\label{fig3}
\end{figure}

\begin{figure}[!ht]
\begin{center}
\begin{tabular}{cc}
(a) & (b)\\
\includegraphics[width=.45\textwidth]{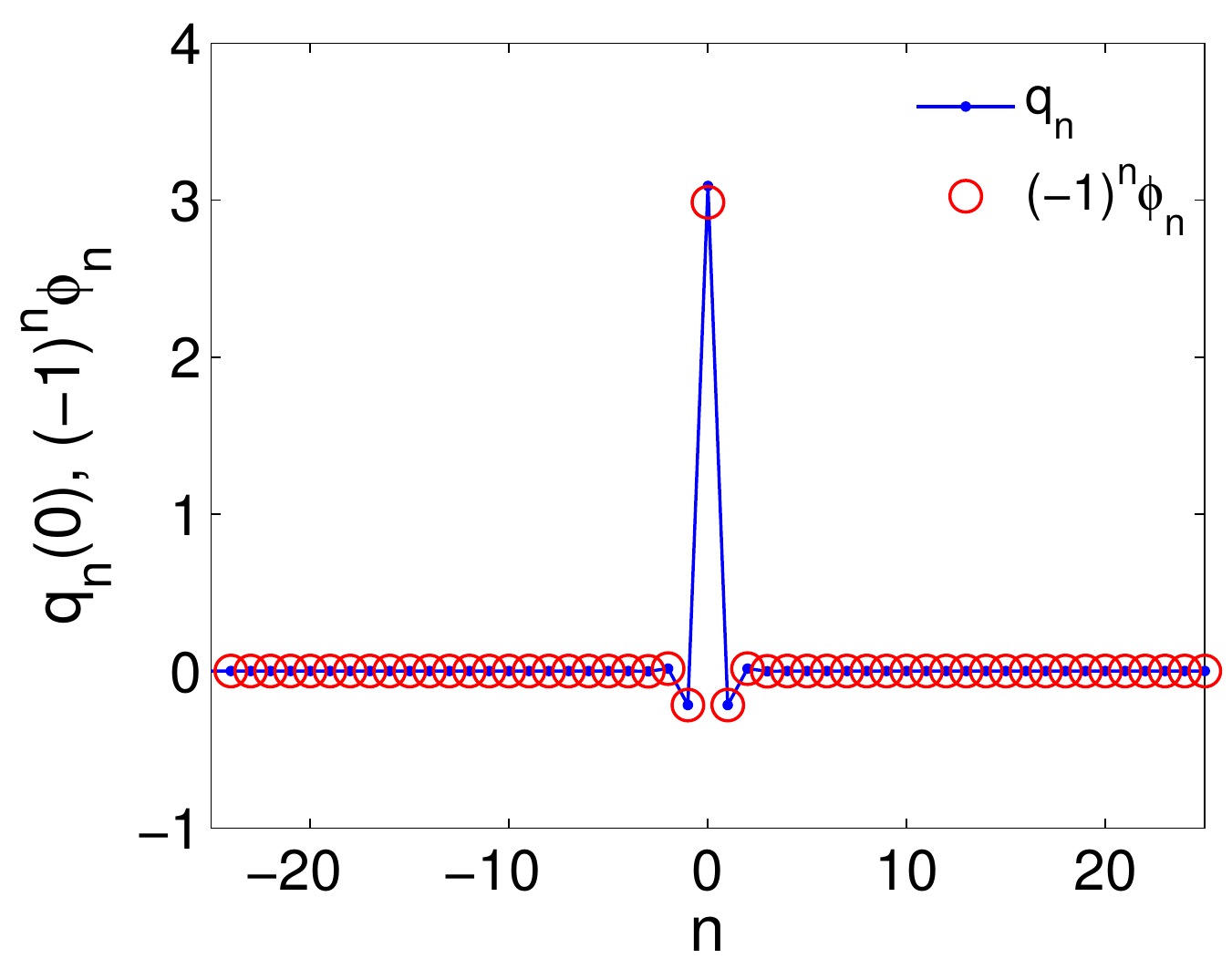} &
\includegraphics[width=.45\textwidth]{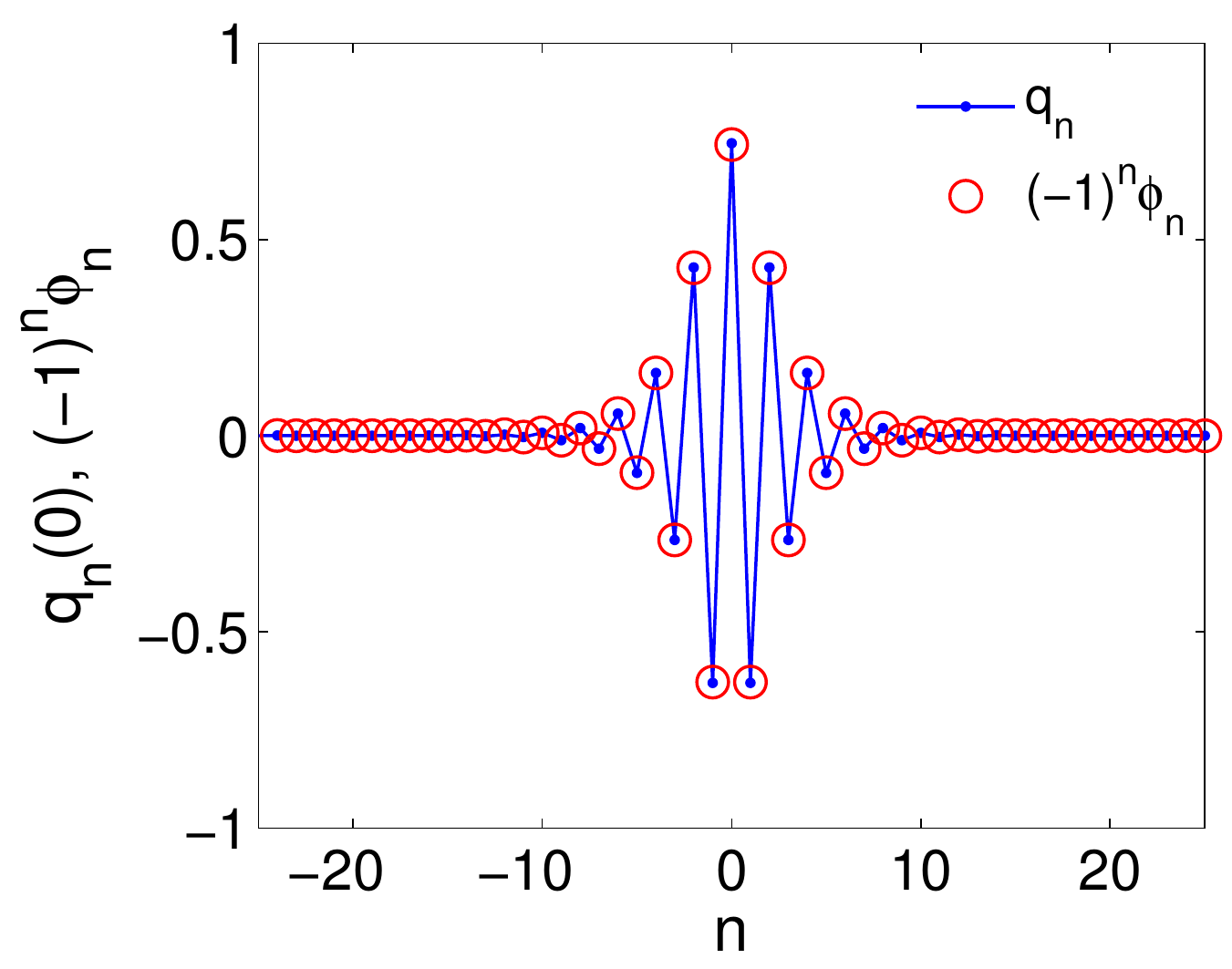}\\
\end{tabular}
\end{center}
\caption{Breather profiles for the sine-Gordon potential in two extreme cases. Panel (a) corresponds to a breather with $\omega_b=3$ and $\kappa=0.5$ whose validity coefficients are $N_\infty=0.0340$ and $N_2=0.0339$. Panel (b) shows a breather with $\omega_b=2$ and $\kappa=0.7$, characterized by $N_\infty=0.0043$ and $N_2=0.0031$. Blue line and dots correspond to the exact profile $\{q_n(0)\}$ and the approximated solution $\{\phi_n\}$ is depicted by red circles.}%
\label{fig4}
\end{figure}

For the hard $\phi^4$ potential, we have a some differences with respect to the sine-Gordon one, as demonstrated in Fig. \ref{fig3}. First of all, the validity coefficients possess, for a given value of the frequency, a monotonically decreasing behaviour so that the maximum value occurs at the anti-continuum limit and becomes zero (i.e. perfect matching) when the breather frequency reaches the phonon band. Secondly, both $N_\infty$ and $N_2$ coefficients exhibit the same behaviour. Figure \ref{fig4} shows the profile of two breathers in a similar fashion to Fig. \ref{fig2}.

\section{Summary}

We have proven the existence of time-periodic spatially localised
solutions
in general infinite nonlinear KG lattice systems utilising the
comparison principle for differential equations.
The time-dependence of
the localised solutions is of uniform, i.e. site-independent, 
functional form being valid for
solutions near the end points of the spectral phonon band. That is,
the frequencies $\omega_b$ of the localised solutions in infinite
KG lattices with soft (hard) on-site potentials
are close to the lowest (highest) normal mode frequency.
To facilitate the existence proof, it is shown that there exist
spatially localised lattice states whose amplitudes of periodic oscillations at each lattice site
are bounded
from above and  below by the oscillating solutions of two auxiliary linear equations.
For soft (hard) on-site potentials the in-phase (out-of-phase) breathers
 oscillate with frequencies slightly  below (above) the lower (upper) value of the
 linear spectrum of phonon frequencies. 
With a  numerical analysis of the features of solutions with frequencies close to the edges of the the linear phonon spectrum 
we have verified that the ensuing 
localised solutions  are of (virtually) uniform time-dependent 
 functional form.

\end{document}